\documentclass[12pt]{article}
\begin{document}
 \def\K{I\!\!K}
\def\N{\mathbb N}
\def\Z{\mathbb Z}
\def\A{\mathbb A}
\def\B{\mathbb B}
\def\C{\mathbb C}
\def\D{\mathbb D}
\def\E{\mathbb E}
\def\R{I\!\!R}
\def\H{I\!\!H}
\def\C{I\!\!\!\!C}
\def\P{I\!\!P}
\def\Q{I\!\!Q}
\def\B{I\!\!B}
\def\F{I\!\!F}
\def\M{I\!\!M}
\def\D{I\!\!D}
\def\P{I\!\!P}
\def\r{\rho}
\def\l{\lambda}
\def\s{\sigma}
\def\b{\beta}
\def\a{\alpha}
\def\d{\delta}
\def\g{\gamma}
\def\z{\zeta}
\def\fp{\hfill \Box \\[2pt]}
\def\e {\epsilon}
\def\proj{\mathop {\rm proj}}
\def\ind{\mathop {\rm ind}}
\def\lim{\mathop {\rm lim}}
\def\id{\mathop {\rm id}}
\def\ker{\mathop {\rm ker}}
\def\max{\mathop {\rm max}}
\def\min{\mathop {\rm min}}
\def\sup{\mathop {\rm sup}}
\def\supp{\mathop {\rm supp}}

\begin{center}
\Large{\bf  AN INTERTWINING OPERATOR FOR THE HARMONIC OSCILLATOR AND THE DIRAC OPERATOR WITH\\ APPLICATION TO THE HEAT AND WAVE KERNELS\\
Ahmedou Yahya ould Mohameden  and  Mohamed Vall Ould Moustapha}
\end{center}

\footnotesize{{\bf Abstract}.   In this article an intertwining operator is constructed which transforms the harmonic oscillator to the Dirac operator  (the first order 
derivative operator). 
  We give also the explicit solutions to the heat and wave equation associated to Dirac operator. As  an application the heat and the wave kernels of the harmonic oscillator are computed.}

\begin{center}
{\bf 1 -- Introduction}
\end{center}
The harmonic oscillators are interesting
and important in their own right and play a fundamental role in the modeling of the quantum fields  and are related to many mathematical and physical problems ($[1], [2] $, $[6] $, $[7]$, $[8]$, $[9]$).
The aim of this paper is to give an intertwining operator $T$ which relate the harmonic oscillator:
 $$L^a=\frac{\partial ^2}{\partial x^2} - a^2 x^2\ \ \ \ \ \ \  a> 0 \eqno(1.1)$$ 
 to the  Dirac operator (first order derivative operator): $$D=\frac{\partial}{\partial X}\eqno(1.2)$$
 We give also
the explicit solutions to the following heat and wave equations associated to Dirac operator:

$$(HD)\qquad \left \{\begin{array}{cc}\partial_t 
U(t,X)=\frac{\partial}{\partial X}U(t,X)&(t,X)\in R^\ast_+\times \R\\ U(0,X)=U_0(X) & U_0\in 
C_0^\infty(\R)\end{array}
\right. $$
$$(WD)\qquad \left \{\begin{array}{cc}\partial_t^2 
V(t,X)=\frac{\partial}{\partial X}V(t,X)&(t,X)\in R^\ast_+\times \R\\ V(0,X)=0 & \partial_t V(0,X)=V_0(X)\in 
C_0^\infty(\R)\end{array}
\right. $$
Note that the Cauchy problem $(HD)$ for the heat equation associated to the Dirac  operator $D$  is nothing but the Cauchy problem for the transport equation.
The wave equation $(WD)$ associated to the Dirac operator is considered in $[3]$ p.$64$ where the unequeness of solution is obtained but there are no known explicit solutions until now.\\
As an application of our intertwining operator, from the heat and wave kernels of the Dirac operator, we give the explicit solutions of the following heat and wave equations associated to the harmonic oscillator:\\
$$(HL^{a})\qquad \left \{\begin{array}{cc}\partial_t
u(t,x)=\left(\frac{\partial^2}{\partial x^2} - a^2 x^2\right)u(t,x)&(t,x)\in R^\ast_+\times \R\\   u(0,x)=u_0(x)& u_0\in 
C_0^\infty(\R)\end{array}
\right. $$ 
 $$(WL^{a})\qquad \left \{\begin{array}{cc}\partial_t^2 
v(t,x)=\left(\frac{\partial^2}{\partial x^2} - a^2 x^2\right)v(t,x)&(t,x)\in \R_+^\ast\times \R\\ v(0,x)=0 & \partial_t v(0,x)=v_0(x), v_0 \in {\cal S}_a(\R)\end{array}
\right.$$.\\
with
$${\cal S}_a(\R)=\left\{\phi : { \cal F} \left[ e^{\frac{-a x^2}{2}}\phi\right](\xi)\in C_0^{\infty} (\R)\right\}$$
Note that the heat kernel for the harmonic oscillator is known for long time $ [1]$:\\ 
$K_{a}(x,x',t)=$$$\sqrt{\frac{a}{2\pi}}\frac{1}{\sqrt{sinh(2at)}}exp\left[-\frac{a}{2}(x^{2}+x'^{2})coth(2at)+\frac{axx'}{sh(2at)}\right]\eqno(1.3)$$
but to our knowledge the method used here and the obtained formula are
new.\\
For the wave kernel of the harmonic oscillator $w_a(t,x,x')$, the following integral representation is given for the wave kernel
at the origin in $[2]  p. 358$:
$$w_1(x,0,t)=\frac{i}{4\pi}\int_{C}\sqrt{\frac{1}{2z \sinh(z)}}e^{\frac{t^{2}}{2z}-\frac{x^{2}c oth(z)}{2}}dz\eqno(1.4)$$
where C is a contour symmetric with respect to the x -axis going throught the origin obtained by a smooth deformation 
of the circle  $C(\frac{1}{2}c,\frac{1}{2}c)$ of center $c/2$ and radius $c/2$ and for $t<\left|x\right|$ the kernel vanishes.\\
Now we recall some facts about  the Fourier transform:\\
for $f\in L^1(\R)$ the Fourier transform of $f$ and its inverse:\\
$$({\cal F}f)(\xi)=\frac{1}{\sqrt{2\pi}}\int_{-\infty}^{+\infty}e^{-i x \xi}f(x) d x\eqno(1.5)$$
$$({\cal F}^{-1}f)(x)=\frac{1}{\sqrt{2\pi}}\int_{-\infty}^{+\infty}e^{i \xi x}f(\xi) d \xi\eqno(1.6) $$
and we have the following formulas:
$${\cal F}f(\alpha\xi)=\frac{1}{\alpha}{\cal F}f(\frac{x}{\alpha})(\xi)\eqno(1.7)$$
$${\cal F}^{-1}[e^{-s\xi^2}](x)=\frac{1}{\sqrt{2s}}e^{-x^2/4s};\ \ \  s>0\ \ \ \eqno(1.8)$$
Recall also the complementary error function $[3]$ p.272
$$Erfc(z)=\int_z^\infty e^{-t^2}dt\eqno(1.9)$$
and in terms of the Tricomi confluent hypergeometric $U(a,c,z)$:
$$Erfc(z)=\frac{1}{2}ze^{-z^2}U(1,3/2,z^2)=\frac{1}{2}e^{-z^2}U(1/2,1/2,z^2)\eqno(1.10)$$
\begin{center}
{\bf 2--The intertwining operator }
\end{center}
{\bf Definition 2.1} An operator $T$ is said to be an intertwining operator if it relates operators, $L$ and $D$, by
$$TL=DT\eqno(2.1)$$

{\bf Lemma 2.2} For $a>0$ and $\phi \in {\cal S}_a(\R)$ we have\\
$\left[\frac{\partial ^2}{\partial x^2} - a^2 x^2\right]\phi(x)=$
$$ e^{\frac{a x^2}{2}}{\cal F}^{-1}\left[e^{-\frac{\xi^2}{4 a}-\frac{log|\xi|}{2}}\,(-2 a \xi \frac{\partial}{\partial \xi})\,e^{\frac{\xi^2}{4 a}+\frac{log|\xi|}{2}}{\cal F}\left[ e^{\frac{-a x^2}{2}}\phi\right](\xi)\right
]( x)\eqno(2.2)$$
{\bf Proof:}
Set
$$\phi(x)=e^{\frac{a x^2}{2}}\psi(x)\eqno(2.3)$$
we get
$$e^{\frac{-a x^2}{2}}\left[\frac{\partial ^2}{\partial x^2} - a^2 x^2\right]e^{\frac{a x^2}{2}}\psi(x)=\left[\frac{\partial ^2}{\partial x^2}+2 a x\frac{\partial}{\partial x}+a\right] \psi (x)\eqno(2.4)$$
take the Fourier transform
$${\cal F}\left[e^{\frac{-a x^2}{2}}\left[\frac{\partial ^2}{\partial x^2} - a^2 x^2\right]e^{\frac{a x^2}{2}}\psi(x)\right](\xi)
=\left[-\xi^2-2 a \xi\frac{\partial}{\partial \xi}-a\right]({\cal F \psi})(\xi)\eqno(2.5)$$
set
$$({\cal F}\psi)(\xi)=e^{-\frac{\xi^2}{4 a}-\frac{log|\xi|}{2}}w(\xi)\eqno(2.6)$$
we obtain
$$e^{\frac{\xi^2}{4 a}+\frac{log|\xi|}{2}}{\cal F}\left[e^{-\frac{a x^2}{2}}\left[\frac{\partial ^2}{\partial x^2} - a^2 x^2\right ] e^{\frac{a x^2}{2}}\psi(x)\right](\xi)
=-2 a \xi\frac{\partial w(\xi)}{\partial \xi} \eqno(2.7)$$
using  $(2.6))$ we have \\
$-2 a \xi\frac{\partial w(\xi)}{\partial \xi}=$
$$e^{\frac{\xi^2}{4 a}+\frac{log|\xi|}{2}}{\cal F}\left[e^{-\frac{a x^2}{2}}\left[\frac{\partial ^2}{\partial x^2} - a^2 x^2\right]e^{\frac{a x^2}{2}}{\cal F}^{-1}[e^{-\frac{\xi^2}{4 a}-\frac{log|\xi|}{2}} w(\xi)](x) \right](\xi)
\eqno(2.8) $$
{\bf Proposition 2.3} If  $f\in {\cal S}(\R)$ and set 
$$(T\phi)(\xi)_{|\xi|=\exp{(-2aX)}}=e^{\frac{\xi^2}{4 a}+\frac{log|\xi|}{2}}{\cal F}\left[ e^{\frac{-a x^2}{2}}\phi\right](\xi)_{|\xi|=\exp{(-2aX)}} \eqno(2.9)$$
Then the operator $T$ is an intertwining operator for the harmonic oscillator $L^a$ given in $(1.1)$ and  the Dirac opetrator $D$ in $(1.2)$:
$$TL^a=DT\eqno(2.10)$$
{\bf Proof} The formula $(2.10)$ is a consequence of the lemma $2.2$
\begin{center}
{\bf 3--Heat and wave kernels for the Dirac operator}
\end{center}
{\bf Proposition 3.1} The Cauchy problem $(HD)$ for the heat equation associated to the Dirac operator has the unique
solution guiven by
$$U(t,X)=U_0(t+X)\eqno (3.1)$$
{\bf Proof: } The Cauchy problem for the heat equation associted to the Dirac operator $D$ reduced to the Cauchy problem
for the classical transport eqution and  its unique solution  is given by $(3.1)$.\\

{\bf Proposition 3.2 } The function $$W(t,X,X')=\frac{t}{\sqrt{4\pi |X-X'|}}\exp{\left(
\frac{-t^{2}}{4|X-X'|}\right)}U\left(1,\frac{3}{2},\frac{t^{2}}{4|X-X'|}\right)\eqno(3.2)$$
satisfies the wave equation associated Dirac operator $D$.\\
and we have
$$W_a(t,X,X')=\frac{2}{\sqrt{\pi}}Erfc\left(\frac{t}{\sqrt{4|X-X'|}}\right)\eqno(3.2)'$$
where $Erfc(x)$ is the complementary error function in $(1.9)$.\\

{\bf Proof}:
Set $z=-\frac{t^2}{4(X-X')}$ we get
$\frac{\partial }{\partial t}=-\frac{2 t}{4(X-X')}\frac{\partial}{\partial z}$,\\
$\frac{\partial^2 }{\partial t^2}=\frac{ t^2}{4(X-X')^2}\frac{\partial^2}{\partial z^2}-\frac{2 }{4(X-X')}\frac{\partial}{\partial z}$,
$\frac{\partial }{\partial X}=\frac{ t^2}{4(X-X')^2}\frac{\partial}{\partial z}$.\\
$$\frac{\partial }{\partial X}\phi(t,X)-\frac{\partial^2}{\partial t^2}\phi(t,X)=\frac{1}{X-X'}\left\{z\frac{\partial}{\partial z^2}+(1/2-z)\frac{\partial}{\partial z}\right\}=0\eqno(3.3)$$
the wave equation associated to the first order derivative operator is equivalent to the equation of confluent hypergeometric type $[5]$ $p.268$
$$z\varphi''(z)+\left(1/2-z\right)\varphi'(z)=0\eqno(3.4)$$
and an appropriate solution $[5]$
$p.270$ is
$z^{1/2}\exp{(z)}U\left(1,3/2,-z\right)$.\\

{\bf Theorem 3.3}
The Cauchy problem $(WD)$ for the wave equation associated to the dirac operator has the unique solution given by:\\
$$V(t,X) = \int_{|X-X'|<~\frac{t}{2}}W\left(X,X'\right)V_0(X')dX'\eqno(3.5)$$
where $W(t,X,X')$ is given by $(3.2)$ and $(3.2)'$.\\

{\bf Proof }
In view of the proposition 3.2  it remain to show the limit conditions in $(WD)$. And for this 
set $X'=X+\frac{t}{2}s$ 
we have
$$V(t,X)=C t^{3/2}\int_{-1}^{1}\frac{V_0(X+s\frac{t}{2})}{\sqrt{|s|}}\exp{\left(-\frac{t}{2|s|}\right)}U\left(1,3/2,\frac{t}{2|s|}\right)ds\eqno(3.6)$$
where $C= 2^{-1/2} \frac{1}{\sqrt{4\pi}}$.
By the formula giving the first derivative of the Lommel confluent hypergeometric function $[4]$ $p.265$
$$\frac{d}{d z}U(a,c, z)=-a U(a+1,c+1,z)\eqno(3.7)$$
we obtain the limit conditions using the following behavior of the  degenerate confluent hypergeometric function $U (a,c,z)$, $[5]$ p.$288-289$
For $z\rightarrow 0$:
$$U(a,c,z)=(\Gamma(c-1)/\Gamma(a))z^{1-c}+ O(1) , 1<\Re c <2\eqno(3.8) $$
$$U(a,c,z)=(\Gamma(c-1)/\Gamma(a))z^{1-c}+ O(|z|^{\Re c-2}) , \Re c\geq 2, c\neq 2\eqno(3.9)$$
\begin{center}
{\bf 4--Heat and wave kernels for the Harmonic oscillator}
\end{center}

The Cauchy problem $(HL^a)$ for heat equation associated to the Harmonic oscillator for $u_0\in L^2(\R)$ has the unique 
solution $u$ belonging to $C^0\left([0,\infty[,L^2(\R)\right)$ see $[6]$ more precisely there exists a semigroup $(S_t)_{t\geq 0}$ of $L^2(\R)$-contractions such that for all $t> 0$, 
$u(t,.)=S_tu_0$, the explicit solution given by
$$u(t,x)=\int_{-\infty}^{+\infty}K_{a}(t,x,x')u_0(x')dx'$$
where the kernel $K_a(t,x,x')$ is given by 
the Mehler's formula $(1.3)$\\

{\bf Theorem 4.1} If $T$ is the intertwining operator given by $(2.9)$ and if  $L^a$ and $ D$ are the harmonic oscillator and the Dirac operator given respectively by $(1.1)$ and $(1.2)$ then we have:
$$e^{tL^a}u_0=T^{-1}\left\{e^{tD}(Tu_0)\right\}\eqno(4.1)$$
$$\frac{\sin t\sqrt{L^a}}{\sqrt{L^a}}v_0=T^{-1}\left\{\frac{\sin t\sqrt{D}}{\sqrt{D}}(Tv_0)\right\}\eqno(4.2)$$
where $e^{tA}$ and  $\frac{\sin t\sqrt{A}}{\sqrt{A}}$ are respectively the heat and the wave kernels for the operator $A$.\\
{\bf Proof}
The formula $(4.1)$  is a consequence of
the formulas $(2.10)$ and the Cauchy problems $(HD)$ and $(HL^a)$.\\
 The formula $(4.2)$  is a consequence of
the formulas $(2.10)$ and the Cauchy problems $(WL^a)$ and $(WD)$.\\

{\bf Theorem 4.2} 
The Cauchy problem $(HL^a)$ for the heat equation associted to the harmonic oscillator has the unique solution given by:\\
$$u(t,x)=\int_{-\infty}^{+\infty}H_{a}(t,x,x') u_0(x') dx'\eqno(4.3)$$
where\\
$H_{a}(t,x,x')=$$$a\sqrt{\frac{2}{\pi}}\left(e^{2at}-e^{-2at}\right)^{-1/2}\exp\left\{\frac{(e^{at}x-e^{-at}x')^2}{e^{2at}-e^{-2at}}+\frac{a}{2}(x^2-x'^2)\right\}\eqno(4.4)$$
{\bf Proof}: Using $(4.1)$ of theorem $4.1$ we have
$$u(t,x)=e^{-at}e^{ax^2/2}{\cal F}^{-1}[e^{-(1-e^{-4at})\xi^2/4a} {\cal F}\left(e^{-ax^2/2}u_0\right)(\xi e^{-2at})](x)
\eqno(4.5)$$
$u(t,x)=\frac{1}{\sqrt{2\pi}}e^{-at}e^{ax^2/2}\times$
$${\cal F}^{-1}[e^{-(1-e^{-4at})\xi^2/4a}]\ast {\cal F}^{-1}[{\cal F}\left(e^{-ax^2/2}u_0\right)(\xi e^{-2at})](x)
\eqno(4.6)$$
using $(1.7)$ and $(1.8)$ we can write
$$u(t,x)=\frac{e^{ax^2/2}}{\sqrt{\pi}}[\frac{\sqrt{a}}{\sqrt{1-e^{-4at}}}e^{-\frac{a x^2}{1-e^{-4at}}}]\ast e^{2at}u_0(xe^{2at})e^{-(a/2)x^2e^{4at}}\eqno(4.7)$$
$$=\frac{e^{at+ax^2/2}}{\sqrt{\pi}}\frac{\sqrt{a}}{\sqrt{1-e^{-4at}}}\int_{-\infty}^{+\infty}u_0(x'e^{2at})e^{-(a/2)x'^2e^{4at}}e^{\frac{-a(x-x')^2}{1-e^{-4at}}}dx'\eqno(4.8)$$
set $x''=x'e^{2at}$ in $(4.8)$ we get the formula $(4.3)$.\\

{\bf Remark 4.3}: The formula in
 $(4.3)$ for the heat kernel associated to the harmonic ocillator agree with that given by $(1.3)$
and the proof is left to the reader.\\

{\bf Theorem 4.4}
The Cauchy problem for the wave equation associated to the harmonic oscillator $(WH)$ has the unique solution given by:\\
$$v(t,x)=-\frac{1}{a\sqrt{\pi}}e^{ax^2/2}{\cal F}^{-1}[\frac{e^{-\xi^2/4a}}{\sqrt{\xi}}\int_{|\ln\xi/\xi'|<at}Erfc\left(\frac{\sqrt{a}t}{\sqrt{2|\ln\xi/\xi'|}}\right)\times$$
 $$\frac{e^{\xi'^2/4a}}{\sqrt{\xi'}}{\cal F}(e^{ax'^2/2}v_0)(\xi')d\xi'](x)\eqno(4.9)$$

{\bf Proof }: The formula $(4.8)$ follows from the theorem $(4.1)$ and the Fubini theorem  using the formulas $(3.2)$ and $(3.2)'$ and
the fact that
$$|Erfc(z)|\leq \sqrt{\pi}$$

We finish this section by the folowing corollary:\\

{\bf Corollary 4.5} The heat and wave kernels for the cursin operator
$$M=\frac{\partial^{2}}{\partial x^{2}}+x^{2}\frac{\partial^{2}}{\partial y^{2}}$$
are given respectively by:
$$~\widetilde{H}_a(t,x,y,x',y',t)~=~\frac{1}{2\pi}\int^{+\infty}_{-\infty}e^{i(y-y')a}H_a(t,x,x') da$$
$$~\widetilde{w}_a(t,x,y,x',y',t)~=~\frac{1}{2\pi}\int^{+\infty}_{-\infty}e^{i(y-y')a}w_a(t,x,x') da$$

\begin{center}{\bf
5--Directions for further studies}
\end{center}

We suggest here a certain number of open related problems connected to this paper.
we are intersted in the heat and wave equations for the harmonic oscillator with an inverse square potential
$$(HL^{a})'\qquad \left \{\begin{array}{cc}\partial_t
u(t,x)=\left(\frac{\partial^2}{\partial x^2} - a^2 x^2-\frac{b^2}{x^2}\right)u(t,x)&(t,x)\in R\times \R\\   u(0,x)=u_0(x)& u_0\in 
C_0^\infty(\R)\end{array}
\right.$$ 
 $$(WL^{a})'\qquad \left \{\begin{array}{cc}\partial_t^2 
u(t,x)=\left(\frac{\partial^2}{\partial x^2} - a^2 x^2-\frac{b^2}{x^2}\right)u(t,x)&(t,x)\in R_+^\ast\times \R\\ u(0,x)=0 & \partial_t u(0,x)=u_0(x)\in 
C_0^\infty(\R)\end{array}
\right. $$.\\
Another possible extension is to consider the heat and wave equation associated to the power of the harmonic oscillator
$$(HL^{a})''\qquad \left \{\begin{array}{cc}\partial_t
u(t,x)=\left(\frac{\partial^2}{\partial x^2} - a^2 x^2\right)^su(t,x)&(t,x)\in R_+^\ast\times \R\\   u(0,x)=u_0(x)& u_0\in 
C_0^\infty(\R)\end{array}
\right. $$ 
 $$(WL^{a})''\qquad \left \{\begin{array}{cc}\partial_t^2 
u(t,x)=\left(\frac{\partial^2}{\partial x^2} - a^2 x^2\right)^su(t,x)&(t,x)\in R_+^\ast\times \R\\ u(0,x)=0 & \partial_t u(0,x)=u_0(x)\in 
C_0^\infty(\R)\end{array}
\right. $$.\\
Finally, we suggest a problems in direction of the non linear heat and wave equations for the harmonic oscillator
and to look for global solution and a possible  blow up in finite times.
\begin{center}\bf
References
\end{center}
${\bf [1]-}$ Berline, N.,Getzler, E., Vergne, M. Heat kernels and dirac operator Springer Verlag 2004.\\
${\bf [2]-}$Greiner P. C. , Daniel Holocman, and Yakar Kannai, Wave kernels related to the second order operator,
Duke Math.J.vol. 114,  $329-387 (2002).$\\
{\bf [3]- }Guelfand, I.M., and Chilov, C. E. Les distributions Tome 3 Th\'eorie des \'equations diff\'erentielles, Dunod 1968\\
${\bf [4]}$-Lebedev, N., N. ; Special Functions
and their applications ; Dover Publications INC New York 1972.\\
{\bf [5]-}Magnus, W. Oberhittenger, F. and Soni R. P.  Formulas and Theorems for the special functions of Mathematical physics,Springer-Verlog New-York 1966.\\
{\bf [6]-} Pazy, A. Semigroups of linear operators and applications to partial differential equations Applied
mathematical sciences $44$ Springer Verlag $1983$.\\
${\bf [7]-}$Sheng-Ya Feng,A note on heat kernels of generalized Hermite operators, Tawanese Journal of Math.  Vol 15,$N^{o}.5,
2035 - 2041 (2011)$.\\
{\bf [8]-}  Read ,S. and Simon B., Methods of modern mathematical physics $II$, Fourier analysis Self-adjointness, Academic  Press, New York - London 1997.\\
{\bf [9]-}Thangavelu Hermite and Laguerre Semigroups some recent developpent Technical Report $N^{o}2006/7$, March 26,2006.\\

\begin{flushleft}
Universit\'e Gaston Berger de Saint-Louis B.P: 234. S\'en\'egal.\\
E-mail adress: ahmeddou2011@yahoo.fr\\
Universit\'e de Nouakchott\\
Facult\'e des sciences et techniques\\
B.P: 5026, Nouakchott-Mauritanie.\\
E-mail adresse: khames@univ-nkc.mr\\
\end{flushleft}
\end{document}